\documentclass[conference]{IEEEtran}
\IEEEoverridecommandlockouts

\usepackage{cite}
\usepackage{url}

\usepackage{amsthm}
\usepackage{mathrsfs}
\usepackage{color}
\usepackage{mathtools,amssymb,nccmath}
\usepackage{bbm}
\usepackage{enumitem}
\usepackage{euscript}       
\usepackage{balance}

\setitemize{labelindent=0.4em,labelsep=0.2cm,leftmargin=0.85cm,topsep=2pt, partopsep=2pt,itemsep=2pt,parsep=2pt}

\theoremstyle{plain}

\newcommand{\ind}[1]{\mathbbm{1}_{\{#1\}}}   
\newcommand{\Exp}{{\mathsf{E}}}
\newcommand{\f}{{\mathsf{f}}}
\newcommand{\g}{{\mathsf{g}}}
\newcommand{\p}{{\mathsf{p}}}
\newcommand{\sfu}{{\mathsf{u}}}

\newcommand{\C}{\mathsf{C}}
\newcommand{\G}{\mathsf{G}}
\newcommand{\D}{\mathsf{D}}
\newcommand{\cC}{\mathscr{C}}
\newcommand{\T}{\mathsf{T}}
\newcommand{\K}{\mathsf{K}}

\newcommand{\cN}{\mathcal{N}}
\newcommand{\cD}{\mathcal{D}}

\newcommand{\cX}{\EuScript{X}}
\newcommand{\cY}{\EuScript{Y}}
\newcommand{\cZ}{\EuScript{Z}}

\begin{document}

\title{Data-Driven Parameter Estimation}

\author{\IEEEauthorblockN{George V. Moustakides}
\IEEEauthorblockA{\textit{Department of Electrical and Computer Engineering} \\
\textit{University of Patras}\\
Patras, GREECE \\
moustaki@upatras.gr}
}
%

\maketitle

\begin{abstract}
Optimum parameter estimation methods require knowledge of a parametric probability density that statistically describes the available observations. In this work we examine Bayesian and non-Bayesian parameter estimation problems under a data-driven formulation where the necessary parametric probability density is replaced by available data. We present various data-driven versions that either result in neural network approximations of the optimum estimators or in well defined optimization problems that can be solved numerically. In particular, for the data-driven equivalent of non-Bayesian estimation we end up with optimization problems similar to the ones encountered for the design of generative networks.
\end{abstract}


\IEEEpeerreviewmaketitle

\section{Introduction}\label{sec:1}

\IEEEPARstart{T}{he} theory of Detection and Estimation constitutes a major background knowledge in Engineering and Statistics. The corresponding methodologies find application in numerous scientific problems and either provide the actual solution or serve as a starting point for developing techniques that are practically implementable. It is remarkable that with very introductory knowledge of Probability Theory one can derive optimum Detection and Parameter Estimation methods \cite{poor,moulin}. In parameter estimation, common denominator in all the optimum approaches is the key assumption that we have a complete statistical description in the form of a joint probability density functions of the observations and the parameters to be estimated (Bayesian) or the observations given the parameters to be estimated (non-Bayesian).

Despite the availability of several popular classes of probability densities, these statistical models tend to fail dramatically when they are used to capture the statistical behavior of modern datasets. The reason is that nowadays data are mostly images or videos enjoying a more structured form which cannot be adequately explained by the usual classes of probability density families (e.g.~Gaussian). It is therefore clear that it is necessary to develop techniques that do not rely on specific density models. 

In most applications there exist sufficient amount of prior data that can be used for \textit{training}, consequently it would be interesting to attempt to develop detection and estimation methods that are \textit{data-driven}, namely do not require exact (or partial) knowledge of probability densities and therefore rely solely on data. Such techniques were developed in \cite{moustakides1} for several versions of the binary hypothesis testing problem based on the direct estimation of the \textit{likelihood ratio} of the two unknown densities which, as we know, is a sufficient statistic for the detection problem.
Similar developments for parameter estimation, to our knowledge, do not seem to exist in any systematic way. Of course it is possible to find pure data-driven estimators for specific estimation problems as for example the estimate of a location parameter but there is no method that provides an answer for a general class of problems. It is this gap we attempt to fill, at least to some extend, developing techniques that are applicable to classes of parameter estimation problems.

%
Our paper is organized as follows: Section\,\ref{sec:1} contains the Introduction. In Section\,\ref{sec:2} we consider data-driven versions of the parameter estimation problem for the Bayesian approach. Section\,\ref{sec:3} constitutes the most important part of our work. We focus on parametric density families generated through parametric transformations and consider the estimation problem under a non-Bayesian framework. In Section\,\ref{sec:4} we apply our methodology to the estimation of a simple data translation problem and evaluate the effectiveness of our idea and how it compares to already existing data-driven methods. 

\section{Data-Driven Bayesian Estimation}\label{sec:2}
In classical Bayesian parameter estimation we assume that a random vector $\cX$ follows the conditional parametric probability density $\f(X|\theta)$, where $\theta$ is the parameter vector that we like to estimate from realizations of $\cX$. Vector $\theta$ is also considered a realization of a random vector $\vartheta$ for which we assume knowledge of a prior density $\p(\theta)$. Combining the two densities we conclude that $\f(X,\theta)=\f(X|\theta)\p(\theta)$ is the joint density of the random \textit{pair} $(\cX,\vartheta)$. Clearly knowing $\f(X,\theta)$ is equivalent to knowing the two densities $f(X|\theta),\p(\theta)$. We recall that an \textit{estimator} of $\theta$ is \textit{any} deterministic vector function $\hat{\theta}(X)$ that has, of course, the same size as $\theta$.

In order to produce the optimum estimator, according to classical Bayesian theory \cite{poor,moulin}, we need to select a cost function $\C(U,\theta)$ and define the average cost
\begin{equation}
\!\cC(\hat{\theta})\!=\!\Exp_{\cX,\vartheta}\big[\C\big(\hat{\theta}(\cX),\vartheta\big)\big]\!=\!\text{\footnotesize$\iint$} \C\big(\hat{\theta}(X),\theta\big)\f(X,\theta)dX d\theta,\!\!\!
\label{eq:2.1}
\end{equation}
where $\Exp_{\cZ}[\cdot]$ denotes expectation with respect to $\cZ$. The average cost must be minimized over the vector function $\hat{\theta}(X)$ in order to produce the optimum estimator $\hat{\theta}_o(X)$, that is,
\begin{equation}
\hat{\theta}_o(X)=\text{arg}\min_{\hat{\theta}(X)}\cC(\hat{\theta})=\text{arg}\min_{\hat{\theta}(X)}\Exp_{\cX,\vartheta}\big[\C\big(\hat{\theta}(\cX),\vartheta\big)\big].
\label{eq:2.2}
\end{equation}
From \cite{poor,moulin} we also know that if we define the function
\begin{multline}
\G(U,X)=\Exp_{\vartheta}\big[\C\big(U,\vartheta\big)|X\big]={\text{\footnotesize$\int$}} \C\big(U,\theta\big)\f(\theta|X)d\theta\\
=\frac{\int \C(U,\theta)\f(X|\theta)\p(\theta)d\theta}{\int\f(X|\theta)\p(\theta)d\theta}=\frac{\Exp_{\vartheta}\big[\C(U,\vartheta)\f(X|\vartheta)\big]}{\Exp_{\vartheta}\big[\f(X|\vartheta)\big]}
\label{eq:2.3}
\end{multline}
where $\f(\theta|X)$ is the \textit{posterior} probability density of $\vartheta$ given $X$ then, using \eqref{eq:2.3} it is also possible to recover the optimum estimator from the following optimization
\begin{equation}
\hat{\theta}_o(X)\!=\!\text{arg}\min_{U}\G(U,X)\!=\!\text{arg}\min_{U}\Exp_{\vartheta}\big[\C(U,\vartheta)\f(X|\vartheta)\big].
\label{eq:2.4}
\end{equation}
We will use \eqref{eq:2.2} and \eqref{eq:2.4} in order to derive data-driven versions of the optimum estimators. We distinguish the following two cases.
\vfill
\subsection{Unknown $\f(X|\theta)$ and $\p(\theta)$}\label{ssec:2.A}
This is the simplest and most straightforward version. The knowledge of the two densities or, equivalently, the knowledge of the joint density $\f(X,\theta)$ is replaced by the availability of realizations of the pair $(\cX,\vartheta)$ suggesting that we must have a collection of pairs $\{(X_1,\theta_1),\ldots,(X_n,\theta_n)\}$. We emphasize that we need realizations of the \textit{pair} which is representative of the random relationship that exists between $\cX$ and $\vartheta$ and expressed through the joint density. If instead we only have two unrelated (i.e.~independent) sets $\{X_1,\ldots,X_n\}$ and $\{\theta_1,\ldots,\theta_n\}$ then this information is clearly not sufficient to capture the random connection between $\cX$ and $\vartheta$.

Since our estimator is a vector function $\hat{\theta}(X)$ that we like to optimize as described above, we can limit our search within a class of vector functions $\sfu(X,\alpha)$ as for example the class of \textit{neural networks} with $\alpha$ denoting the \textit{network parameters}. Replacing $\hat{\theta}(X)$ in \eqref{eq:2.1} with $\sfu(X,\alpha)$ defines a new average cost that depends only on the network parameters $\alpha$
\begin{equation}
\tilde{\cC}(\alpha)=\Exp_{\cX,\vartheta}\big[\C\big(\sfu(\cX,\alpha),\vartheta\big)\big].
\label{eq:2.5}
\end{equation}
As in \eqref{eq:2.2}, we would like to minimize this criterion over $\alpha$ in order to optimize $\sfu(X,\alpha)$, that is,
\begin{equation}
\alpha_o=\text{arg}\min_{\alpha}\tilde{\cC}(\alpha)=\text{arg}\min_{\alpha}\Exp_{\cX,\vartheta}\big[\C\big(\sfu(\cX,\alpha),\vartheta\big)\big].
\label{eq:2.6}
\end{equation}
The optimization in \eqref{eq:2.6} is in the classical form that accepts computation of $\alpha_o$ using the stochastic gradient descent algorithm applied to the training data $\{(X_1,\theta_1),\ldots,(X_n,\theta_n)\}$. Specifically we have
\begin{multline}
\alpha_t=\alpha_{t-1}-\mu\nabla_{\!\alpha}\C\big(\sfu(X_t,\alpha_{t-1}),\theta_t\big)\\[2pt]
=\alpha_{t-1}-\mu\big[\mathbb{J}_{\alpha}\sfu(X_t,\alpha_{t-1})\big]^\intercal\big[\nabla_{\!U}\C\big(\sfu(X_t,\alpha_{t-1}),\theta_t\big)\big],
\label{eq:2.7}
\end{multline}
where $\mathbb{J}_{\alpha}\sfu(X,\alpha)$ denotes the Jacobian of $\sfu(X,\alpha)$ with respect to $\alpha$ and $\nabla_{\!U}\C\big(U,\theta)$ the gradient of $\C\big(U,\theta)$ with respect to $U$.
We recall that $\mu>0$ is the step size (learning rate) and that in each iteration $t$ we employ a pair $(X_t,\theta_t)$ from the available training data. If the data are exhausted before we reach convergence then we can reuse them after, possibly, applying a random permutation. Alternatively, we could approximate expectation with sample means and replace \eqref{eq:2.6} with{{\parfillskip0pt\par}}
\newpage
\null\vskip-0.5cm
$$ 
\alpha_o=\text{arg}\min_{\alpha}\frac{1}{n}\sum_{i=1}^n\C\big(\sfu(X_i,\alpha),\theta_i\big),
$$ 
that accepts a gradient descent iterative solution of the form
\begin{multline*}
\alpha_t=\alpha_{t-1}\\[-2pt]
-\mu\frac{1}{n}\sum_{i=1}^n\big[\mathbb{J}_{\alpha}\sfu(X_i,\alpha_{t-1})\big]^\intercal\big[\nabla_{\!U}\C\big(\sfu(X_i,\alpha_{t-1}),\theta_i\big)\big].
\end{multline*}
If either of the two algorithms converges to $\hat{\alpha}_o$ and if this limit does not correspond to some local minimum, then we expect that we can approximate the optimum estimator as follows
\begin{equation}
\hat{\theta}_o(X)\approx\sfu(X,\hat{\alpha}_o).
\label{eq:2.8}
\end{equation}
In other words, we anticipate that the output of $\sfu(X,\hat{\alpha}_o)$ will provide estimates that are close to the estimates of the optimum estimator $\hat{\theta}_o(X)$. 

From the above it is clear that in this case we compute a mathematical formula for the estimator in the form, for example, of a neural network. We must however point out that the application of iterative solvers based on gradients is possible only if we can find the gradient of $\C(U,\theta)$ with respect to $U$. This is clearly the case in the MMSE criterion where $\C(U,\theta)=\|U-\theta\|^2$ or the MAE criterion with $\C(U,\theta)=\|U-\theta\|_1$. Unfortunately the same observation does not apply in the case of the popular MAP estimator where $\C(U,\theta)=\ind{\|U-\theta\|>\delta}$ with $0<\delta\ll1$ and $\mathbbm{1}_A$ denoting the indicator function of the set $A$. This is because the indicator takes values 1 or 0 with derivative that is either 0 at non-boundary points or $\infty$ at boundary points. Of course it is always possible to approximate $\ind{\|U-\theta\|>\delta}$ with some smooth and differentiable functions but we will still experience computational problems because $\delta$ must be selected very small suggesting that in \eqref{eq:2.7} we will rarely observe gradients that are not close to 0. This will clearly affect the convergence speed of the iterations producing excessive convergence delays. Consequently the MAP estimator requires a substantially different approach for which, unfortunately, we have no meaningful answer at the moment.

\subsection{Known $\f(X|\theta)$ and unknown $\p(\theta)$}\label{ssec:2.B}
Although we argued that we are interested in abandoning the assumption of known probability densities, we would like to consider the case where $\f(X|\theta)$ is known. The reason is that in some applications this assumption is regarded as realistic. The prior $\p(\theta)$ of the random parameter $\vartheta$ on the other hand, as in the previous case, is considered unknown and replaced by the availability of a set of realizations $\{\theta_1,\ldots,\theta_n\}$ that follow $\p(\theta)$. The goal is to employ directly this dataset to obtain Bayesian-like estimates of the desired parameters instead of using it to estimate the prior density $\p(\theta)$ first and then apply the classical Bayesian theory.

Here we are not necessarily targeting the development of a mathematical expression for the estimator since we no longer have realizations of $\cX$ that could be used for training. We recall that we have assumed that we know the functional form of the conditional density $\f(X|\theta)$. 
Suppose now that we are given a realization $X$ of $\cX$ for which we would like to obtain the corresponding estimate of $\theta$. Following \eqref{eq:2.4} we have that
\begin{equation}
\hat{\theta}_o(X)=\text{arg}\min_{U}\Exp_{\vartheta}\big[\C(U,\vartheta)\f(X|\vartheta)\big],
\label{eq:2.9}
\end{equation}
which involves averaging \textit{only} with respect to $\vartheta$ while $\f(X|\theta)$ for given $X$ is a known function of $\theta$. The optimization in \eqref{eq:2.9} is in the standard form that accepts a stochastic gradient descent algorithmic solution of the form
\begin{equation}
U_t=U_{t-1}-\mu\left[\nabla_{\!U}\C(U_{t-1},\theta_t)\right]\f(X|\theta_t).
\label{eq:2.10}
\end{equation}
The limit of the sequence $\{U_t\}$ generated by the iterative procedure in \eqref{eq:2.10} will constitute the desired estimate $\hat{\theta}_o(X)$. Alternatively, one may approximate the expectation in \eqref{eq:2.9} with the sample mean and attempt the minimization
\begin{equation}
\hat{\theta}_o(X)=\text{arg}\min_{U}{\frac{1}{n}\sum_{i=1}^n}\C(U,\theta_i)\f(X|\theta_i)
\label{eq:2.9b}
\end{equation}
which whenever not possible to solve analytically it can give rise to a gradient descent algorithm of the form
\begin{equation}
U_t=U_{t-1}-\mu\frac{1}{n}\sum_{i=1}^n\left[\nabla_{\!U}\C(U_{t-1},\theta_i)\right]\f(X|\theta_i),
\label{eq:2.11}
\end{equation}
every time the data vector $X$ is given.
We observe that both iterations \eqref{eq:2.10} and \eqref{eq:2.11}, for every \textit{given} observation vector $X$,  employ the dataset $\{\theta_1,\ldots,\theta_n\}$ and the knowledge of the density $\f(X|\theta)$ in order to compute \textit{numerically} the desired estimate.

There are cases where it is possible to solve \eqref{eq:2.9b} directly and obtain a mathematical formula for the corresponding estimator. 
When $\C(U,\theta)=\|U-\theta\|^2$, that is, when we are interested in the MMSE, it is easy to verify that this leads to the following estimator function
\begin{equation}
\hat{\theta}_{\rm MMSE}(X)=\frac{\sum_{i=1}^n\theta_i\,\f(X|\theta_i)}{\sum_{i=1}^n\f(X|\theta_i)},
\label{eq:2.12}
\end{equation}
with the right hand side being an approximation of the conditional expectation of $\theta$ given $X$ which is the ideal MMSE estimator. The resulting formula is clearly \textit{not in the form of a neural network}. A closed form solution is also possible in the case of the minimum mean absolute error (MMAE) however, due to lack of space we are not going to present it.

Completing our presentation of the Bayesian-like data-driven estimators we must add that, as in Section\,\ref{ssec:2.A}, iterative (stochastic) gradient descent algorithms are impossible to apply in the case of the MAP estimator because, as before, we cannot compute the gradient of $\C(U,\theta)$ with respect to $U$. 


\section{Data-Driven non-Bayesian Estimation}\label{sec:3}
Let us now examine the far more interesting problem of non-Bayesian parameter estimation. In its classical version we are given a density $\f(X|\theta)$ that contains a parameter vector $\theta$ which is considered \textit{deterministic and unknown}. In other words there exist no prior density that describes its statistical behavior. The most popular means to solve this parameter estimation problem \cite{poor,moulin} is by employing the Maximum Likelihood Estimator (MLE), namely
\begin{equation}
\hat{\theta}(X)=\text{arg}\max_{\theta}\f(X|\theta).
\label{eq:ML}
\end{equation}
This estimator, under general conditions enjoys asymptotic optimality (as the length of $X$ tends to infinity) in the sense that in the limit its error covariance matrix \textit{approaches the Cramer-Rao Lower Bound} (CRLB).

Defining a data-driven version for this estimation problem is not as straightforward as in the Bayesian case. First of all because there is no prior for $\theta$ this immediately translates in the data-driven setup that \textit{there are no realizations} of $\theta$ which could be used for training. Since the idea is to replace probability densities with data sampled from these densities we assume that we have available a set of data $\{X_1,\ldots,X_n\}$ that follow the unknown density $\f(X|\theta)$ \textit{for the same} $\theta$. If we try to solve our problem at this stage, it is possible to employ different data-driven formulations with drastically different answers but without any means to decide which is the most appropriate solution. To be able to proceed we need to impose additional structure on the problem of interest that will allow us to produce a version which makes sense from a practical as well as theoretical point of view.

A possible direction we may follow is to define the parametric density $\f(X|\theta)$ \textit{indirectly}. Let us start with a random vector $\cZ$ which is distributed according to the density $\g(Z)$. Consider now a \textit{deterministic transformation} $\T(Z,\theta)$ that contains the parameter vector $\theta$. By defining with the help of the transformation the new random vector $\cX=\T(\cZ,\theta)$ it is clear that $\cX$ will have a probability density $\f(X|\theta)$ which is a function of $\theta$. We note that the parametric density $\f(X|\theta)$ of course exists but we do not necessarily have its explicit form. We must also point out that the transformation does not have to be one-to-one since the resulting $\cX$ can be of dimension larger than the dimension of $\cZ$ allowing $\cX$ to live on a lower dimensional manifold.
%
Consider now the following parameter estimation problem.
\vskip0.1cm
\noindent\textbf{Non-Bayesian Parameter Estimation Problem:} \textit{Assume a random vector $\cZ$ is distributed according to the density $\g(Z)$. A transformation $\T(Z,\theta)$ with parameters $\theta$ is applied onto $\cZ$ generating the random vector $\cX=\T(\cZ,\theta)$ which follows the unknown density $\f(X|\theta)$. We are given the dataset $\{X_1,\ldots,X_n\}$ comprised of independent realizations of $\cX$ generated with the same $\theta$
and the dataset $\{Z_1,\ldots,Z_m\}$ with independent realizations of $\cZ$ that follow $\g(Z)$. Assuming knowledge of the functional form of the transformation $\T(Z,\theta)$ we would like to estimate the parameter vector $\theta$ that gives rise to the first dataset.}

This is clearly a parameter estimation problem which is purely data-driven since there is no knowledge of any probability density. One might argue that we do not need any densities since from the correspondence $X_i=\T(Z_i,\theta)$ it is a simple exercise to estimate $\theta$ by minimizing some form of distance between the two sides. However, \textit{this is not true} because the two datasets are considered entirely unrelated being sampled independently with no actual correspondence between their samples. The first dataset is simply a representative of $\f(X|\theta)$ (containing the information about $\theta$) while the second is a representative of $\g(Z)$.

With this class of parametric densities generated with the help of parametrized transformations we cannot, of course, capture the generality of the original parameter estimation problem where $\f(X|\theta)$ can be any parametric density. However the proposed class is fairly rich with some classical parameter estimation problems being straightforward examples of the proposed data model. For instance if $\T(Z,\theta)=Z+\theta$ then this corresponds to an unknown translation of the random vector $\cZ$. In fact this simple transformation will be used in our simulation experiments in Section\,\ref{sec:4}. Another well known instance of our setting is a change of scale in each element of $\cZ$ where $\T(Z,\theta)=\theta\odot Z$ with ``$\odot$'' denoting the element-by-element multiplication of the two vectors. Finally, a more challenging version would be $\T(Z,\Theta)=\Theta Z$ where $\Theta$ is an unknown matrix that replaces the parameter vector $\theta$. Clearly, one can come up with more complex examples that are not necessarily linear as the cases we mentioned.

Possible solution to the non-Bayesian parameter estimation problem constitutes the moment matching method proposed in \cite{li} where moments of $\cX$ estimated from the first dataset $\{X_1,\ldots,X_n\}$ are matched to the corresponding moments of $\T(\cZ,\theta)$ using the transformed second dataset $\{\T(Z_1,\theta),\ldots,\T(Z_m,\theta)\}$ thus defining suitable equations. Employing an adequate number of such equations we can solve for the unknown parameters $\theta$. Unfortunately there is an infinite number of moment combinations that could be used to solve the same problem and, more importantly, we recall that classical moment estimation methods are notoriously non-robust hence easily resulting in unsatisfactory performance.

\subsection{Density Matching}
Instead of attempting to match moments we could alternatively select the parameters \textit{to match the two probability densities} of the two datasets $\{X_1,\ldots,X_n\}$ and $\{\T(Z_1,\theta),\ldots,\T(Z_m,\theta)\}$. For density matching it is not necessary to estimate the two densities. For example, it would be sufficient to estimate the \textit{likelihood ratio function} and by properly selecting the parameters $\theta$ to bring this function as close as possible to~1 (perfect match). 

The idea we just mentioned is motivated by the results in \cite{goodfellow} where the notion of Generative Adversarial Networks (GANs) was first introduced. We recall from \cite{goodfellow} that we have a random vector $\cX$ that follows a density $f(X)$ and we are interested in generating realization of $\cX$. This is achieved by first generating realizations of $\cZ$ which follows some density $g(Z)$ and then applying a transformation $\cY=\G(\cZ)$ with $\G(Z)$ known as the ``generator'' function. The generator $\G(Z)$ is designed so that \textit{the density of $\cY$ matches the density of $\cX$}. The ``matching quality'' is evaluated with the help of the ``discriminator'' function $\D(X)$ that tries to differentiate between the ``true'' $\cX$ and the ``synthetic'' $\G(\cZ)$. Matching is achieved when the selected generator makes the discriminator fail in its task to distinguish the statistical behavior of the two random vectors $\cX$ and $\cY$. In \cite{goodfellow} it is proved that the generator/discriminator pair which is capable of achieving the desired matching is the solution to the following min-max (adversarial) problem
\begin{equation}
\min_{\D(X)}\max_{\G(Z)}\big\{\Exp_{\cX}[\log\D(\cX)]+\Exp_{\cZ}\big[\log\big(1-\D\big(\G(\cZ)\big)\big)\big]\big\}.
\label{eq:ideal}
\end{equation}
%

This first original work was followed by a number of alternative methods that appeared in the literature all adopting a similar adversarial setup. We must mention \cite{arjovsky} but also the more general result in \cite{moustakides2,moustakides3}. Regarding the latter approach, the problem in \eqref{eq:ideal} is extended to
\begin{equation}
\min_{\G(Z)}\max_{\D(X)}\big\{\Exp_{\cX}[\phi\big(\D(\cX)\big)]+\Exp_{\cZ}\big[\psi\big(\D\big(\G(\cZ)\big)\big)\big]\big\},
\label{eq:ideal2}
\end{equation}
with the two functions $\phi(z),\psi(z)$ satisfying $\psi'(z)=\rho(z)$,
$\phi'(z)=-\omega^{-1}(z)\rho(z)$,
where ``$\,'\,$'' denotes derivative, $\rho(z)>0$ is a strictly positive function and $\omega(r)$ is a strictly increasing differentiable function defined on $r\in[0,\infty)$ with $\omega^{-1}(z)$ being its inverse function. In \cite{moustakides2,moustakides3} it is then proved that the adversarial problem in \eqref{eq:ideal2} produces a generator/discriminator pair with the generator output $\cY=\G(\cZ)$ matching the \textit{statistical behavior} of $\cX$ (i.e.~its density). 
In \cite{moustakides2} one can find a plethora of pairs following the above rules which are successful in identifying the right generator function.

%

Adversarial approaches when applied to the design of GANs are well known to suffer from convergence instability when implemented iteratively using stochastic gradients. Of course we must also not forget the fact that we are interested in designing a generator and we end up designing also a second function, the discriminator, which becomes useless once the generator is computed. One can find very few generator design techniques that do not need a discriminator function. These methods are \textit{non-adversarial} suggesting that their implementation is going to be far more stable than their adversarial counterparts. We focus on a specific such technique introduced in \cite{moustakides4} which is called \textit{Maximal Correlation} method and consists in solving the following optimization problem with respect to the generator $\G(Z)$
\begin{equation}
\max_{\G(Z)}\frac{\Exp_{\cX,\cZ^1}\big[\K\big(\cX,\G(\cZ^1)\big)\big]\Exp_{\cX,\cZ^2}\big[\K\big(\cX,\G(\cZ^2)\big)\big]}{\Exp_{\cZ^1,\cZ^2}\big[\K\big(\G(\cZ^1),\G(\cZ^2)\big)\big]},
\label{eq:ideal22}
\end{equation}
where $\K(X,Y)$ is a positive definite kernel and $\cZ^1,\cZ^2$ are two independent random vectors with the same density $g(Z)$ while $\cX$ follows $f(X)$. As it is shown in \cite{moustakides4} the generator function that maximizes the correlation in \eqref{eq:ideal22} when used in the transformation $\cY=\G(\cZ)$ the $\cY$ it produces matches $\cX$ in density exactly as in the adversarial problems. Since here we have a single optimization its implementation is far easier. 

The connection of the two approaches in \eqref{eq:ideal} and \eqref{eq:ideal22} to our parameter estimation problem is not difficult to see. In our case we do \textit{not need to identify the generator function $\G(Z)$} since this role is undertaken by the parametric transformation $\T(Z,\theta)$. Focusing on \eqref{eq:ideal22} which we are going to adopt in our simulation experiments 
the optimization problem becomes
%
%
$$ 
\max_{\theta}\frac{\Exp_{\cX,\cZ^1}\big[\K\big(\cX,\T(\cZ^1,\theta)\big)\big]\Exp_{\cX,\cZ^2}\big[\K\big(\cX,\T(\cZ^2,\theta)\big)\big]}{\Exp_{\cZ^1,\cZ^2}\big[\K\big(\T(\cZ^1,\theta),\T(\cZ^2,\theta)\big)\big]},
$$ 
with the optimization over $\G(Z)$ being replaced by the optimization over the parameters $\theta$.
By approximating expectations with sample means gives rise to the data-driven version of the optimization problem that will provide our desired parameter estimates. Specifically we are interested in
\begin{equation}
\max_{\theta}\frac{\cN_1(\theta)\cN_2(\theta)}{\cD(\theta)}
\label{eq:data2}
\end{equation}
where
\begin{align}
\begin{split}
\cN_\ell(\theta)&=\sum_{i=1}^n\sum_{j=1}^{m_\ell}\K\big(X_i,\T(Z_j^\ell,\theta)\big),~\ell=1,2\\
\cD(\theta)&=\sum_{i=1}^{m_1}\sum_{j=1}^{m_2}\K\big(\T(Z_i^1,\theta),\T(Z_j^2,\theta)\big),
\end{split}
\label{eq:data3}
\end{align}
and where we have split the second dataset $\{Z_1,\ldots,Z_m\}$ into two independent parts $\{Z_1^1,\ldots,Z_{m_1}^1\}$ and $\{Z_1^2,\ldots,Z_{m_2}^2\}$.


\section{Experiments}\label{sec:4}
We focus on the non-Bayesian version and in particular the first problem mentioned in Section\,\ref{sec:3} namely the estimation of an unknown translation of the data. For simplicity we limit ourselves to the scalar case.
Let us begin with a density $\g_0(W)$ which is zero mean. We then define $\g(Z)=\g_0(Z-\mu)$ where $\mu$ is an initial unknown mean of $Z$. Next we apply a translation $\theta$ that results in $\f(X|\theta)=\g(X-\theta)=\g_0(X-\mu-\theta)$. The goal is to estimate $\theta$ using the two datasets $\{X_1,\ldots,X_n\}$ and $\{Z_1,\ldots,Z_m\}$ sampled from $\f(X|\theta)$ and $\g(Z)$ respectively. This suggests that we could first estimate $\mu$ using $\{Z_1,\ldots,Z_m\}$ and then $\mu+\theta$ from $\{X_1,\ldots,X_n\}$. The two estimates must then be subtracted $\hat{\theta}=\widehat{\mu+\theta}-\hat{\mu}$ in order to produce the desired estimate of $\theta$. Since each estimate employs a different dataset and since the two datasets are independent we can write
$$ 
\Exp[(\hat{\theta}-\theta)^2]\!=\!\Exp[(\widehat{\mu+\theta}-\mu-\theta)^2]+\Exp[(\hat{\mu}-\mu)^2]
\!\geq\!\frac{1}{\text{FI}}\left(\frac{1}{n}+\frac{1}{m}\right),
$$ 
where we have lower bounded each error power by its corresponding CRLB with the Fisher Information satisfying $\text{FI}=\int\frac{(\g_0'(W))^2}{\g_0(W)}dW$.
No estimator of $\theta$ in the form $\hat{\theta}=\widehat{\mu+\theta}-\hat{\mu}$ can enjoy an error power smaller than the lower bound we have specified. We know \cite{poor,moulin} that this lower bound is attained asymptotically for large $n,m$ by the MLE with the corresponding estimate being
$$
\hat{\theta}_{\rm MLE}\!=\!\text{arg}\max_{\nu}\!\sum_{i=1}^n\log\g_0(X_i-\nu)-\text{arg}\max_{\mu}\!\sum_{i=1}^m\log\g_0(Z_i-\mu).
$$
where $\nu$ replaces the sum $\mu+\theta$.
Of course the previous estimate is not data-driven since it requires knowledge of the density $\g_0(W)$.

The most obvious data-driven estimator of $\theta$ is clearly the one that matches the first moments by combining the two
 sample means, that is
\begin{equation}
\hat{\theta}_{\rm M}=\frac{1}{n}\sum_{i=1}^nX_i-\frac{1}{m}\sum_{i=1}^mZ_i.
\label{eq:MM}
\end{equation}
As before, because of the independence of the two datasets
we can easily show that
$$
\Exp[(\hat{\theta}_{\rm M}-\theta)^2]=\sigma_0^2\left(\frac{1}{n}+\frac{1}{m}\right),
$$
where $\sigma_0^2=\int W^2\g_0(W) dW$ is the corresponding variance. Since sample means are well known to be non-robust one can develop robust alternatives by adopting the approach in \cite{huber}
$$ 
\hat{\theta}_{\rm R}=\text{arg}\min_{\nu}{\sum_{i=1}^n\varphi(X_i-\nu)}-
\text{arg}\min_{\mu}{\sum_{i=1}^m\varphi(Z_i-\mu)},
$$ 
where $\varphi(W)$ is a proper convex function. For example $\varphi(W)=W^2$ results in the estimator in \eqref{eq:MM}, while selecting $\varphi(W)$ to be the Huber function \cite{huber}
$$
\varphi(W)=\left\{\begin{array}{cl}W^2&\text{for}~|W|\leq c\\[2pt]
2c|W|-c^2&\text{for}~|W|>c,\end{array}\right.
$$
constitutes a popular method to robustify the estimates of the two means (location parameters) in \eqref{eq:MM}.
For the proposed maximal correlation method in \eqref{eq:data2},\eqref{eq:data3} we consider the Gaussian kernel $\K(X,Y)=e^{-\frac{\|X-Y\|^2}{h}}$ with $h=1$.

Regarding the sizes of the two datasets, we examine the case $n=m=100$ and the second dataset in the maximal correlation method is split into $m_1=m_2=\frac{m}{2}=50$ samples. Finally for the density $\g_0(W)$ we simulate three cases: 1)~Gaussian with $\g_0(W)=(2\pi)^{-1/2}e^{-W^2/2}$, FI=1, $\sigma_0^2=1$; 2)~Laplace with $\g_0(W)=0.5e^{-|W|}$, FI=1, $\sigma_0^2=2$ and 3)~Cauchy with $\g_0(W)=\frac{1}{\pi}\frac{1}{1+W^2}$, FI=0.5, $\sigma_0^2=\infty$, that exhibit increasing tail fatness. In all three cases we select $\theta=\mu=1$. For the MLE, the moment matching, the Huber robust estimator with $c=1$ and the maximal correlation (proposed) the error power is computed by averaging over 100,000 independent runs.

\begin{table}[!h]
\vskip-0.2cm
\centering
\caption{Error power of translation estimates.}
\label{tab:1}
\vskip-0.1cm
\renewcommand{\arraystretch}{1.4}
\begin{tabular}{lcccl}
&Gaussian&Laplace&Cauchy&\\
CRLB&0.020&0.020&0.040&\\
MLE&0.020&0.023&0.041&\\
Moment Matching&0.020&0.040&$\infty$&Data-driven\\
Huber Estimator&0.022&0.027&0.052&Data-driven\\
Maximal Correlation&0.026&0.026&0.044&Data-driven
\end{tabular}
\vskip-0.2cm
\end{table}
In Table\,\ref{tab:1} for each estimator we present the error power. In the case of Gaussian data we know that the simple moment matching estimator in \eqref{eq:MM} is the same as the MLE and attains the CRLB for every \textit{finite} $n,m$. However performance degrades rapidly as we diverge from Gaussianity and use data from fat-tailed densities. The main observation is that our method is antagonistic to Huber's robust estimator being also very close to the MLE which is not data-driven. At the same time our idea enjoys the advantage of being applicable to \textit{any} transformation $\T(Z,\theta)$ as opposed to Huber's robust approach which is primarily employed for the location parameter problem.


\ifCLASSOPTIONcaptionsoff
  \newpage
\fi

\balance


\begin{thebibliography}{10}

\bibitem{poor} H.V.~Poor, {\em An Introduction to Signal Detection and Estimation}, 2nd Ed, Springer, 1994.

\bibitem{moulin} P.~Moulin, V.V.~Veeravalli, {\em Statistical Inference for Engineers and Data Scientists}, Cambridge, NY, 2019.

\bibitem{moustakides1} G.V.~Moustakides, K.~Basioti, ``Training neural networks for likelihood/density ratio estimation,'' \textit{arXiv:\,1911.00405}, Nov. 2019.

\bibitem{li} C.-L.~Li et al., ``MMD GAN: Towards deeper understanding of moment matching network,'' \textit{arXiv:\,1705.08584}, 2017.

\bibitem{goodfellow} J.~Goodfellow et al., ``Generative adversarial networks,'' \textit{arXiv: 1406. 2661}, 2014.

\bibitem{arjovsky} M.~Arjovsky, S.~Chintala, and L.~Bottou, ``Wasserstein generative adversarial networks,'' \textit{International Conference on Machine Learning}, PMLR, pp.~214--223, 2017.

\bibitem{moustakides2} K.~Basioti, G.V.~Moustakides, ``Designing GANs: A likelihood ratio approach,'' \textit{arXiv:\,2002.00865}, Feb. 2020.

\bibitem{moustakides3} K.~Basioti, G.V.~Moustakides, ``Generative adversarial networks: A likelihood ratio approach,'' \textit{International Joint Conference on Neural Networks}, Shenzhen, China, July 2021. 

\bibitem{moustakides4} K.~Basioti, G.V.~Moustakides, E.Z.~Psarakis, ``Maximal correlation: An alternative criterion for training generative networks,'' \textit{24th European Conference on Artificial Intelligence}, Santiago de Compostela, Spain, June 2020. 

%


\bibitem{huber} P.J~Huber, ``Robust estimation of a location parameter,'' \textit{Ann. Math. Statist.} vol.~35, no.~1, pp.~73--101, 1964.



\end{thebibliography}
\end{document}